\documentclass[preprint]{aastex}
\usepackage{float}
\usepackage{multirow}

\newcommand{\sgr}{\mbox{SGR\,J0501$+$4516~}}
\newcommand{\sgrnos}{\mbox{SGR\,J0501$+$4516}}
\newcommand{\sgrb}{\mbox{SGR\,J1550$-$5418~}}
\newcommand{\sgrbnos}{\mbox{SGR\,J1550$-$5418}}

\shorttitle{An Indepth Study of the Dimmest Bursts from Magnetars}
\shortauthors{Lin Lin}

\begin{document}

\title{Detailed Investigations of the Dimmest Bursts from Two Magnetars, \sgr and \sgrb}

\author{Lin Lin\altaffilmark{1}, Ersin G{\"o}{\v g}{\"u}{\c s}\altaffilmark{1}, Yuki Kaneko\altaffilmark{1}, and Chryssa Kouveliotou\altaffilmark{2}}

\email{lin198361@gmail.com}

\altaffiltext{1}{Sabanc\i~University, Faculty of Engineering and Natural Sciences, Orhanl\i$-$ Tuzla, \.{I}stanbul 34956, Turkey}
\altaffiltext{2}{Science and Technology Office, ZP12, NASA Marshall Space Flight Center, Huntsville, AL 35812, USA}

\begin{abstract}
We applied the Bayesian blocks representation technique to search for the dimmest bursts from two magnetars: we identified 320 events from \sgr using a deep \textit{XMM-Newton} observation and 404 bursts from \sgrb using two \textit{Swift}/XRT pointings. The fluence level of our sample for both sources are about $1-2$ orders of magnitude lower than earlier studies. We systematically investigated morphological characteristics and duration distributions of these bursts, as these properties are directly obtained from their Bayesian blocks profiles. We also studied the spectral behavior of the dimmest bursts, which were grouped based on the morphological types and fluences. Our results helped us further differentiate the spectral nature of very dim bursts from that of the persistent emission, both fitted with physically motivated continuum emission models. Moreover, we generated the differential burst fluence distribution for these two magnetars in the lowest fluence regime. 

\end{abstract}


\section{Introduction}

The most extreme magnetic fields in the Universe have been detected from a group of isolated neutron stars, known as magnetars. About 24 magnetars have been identified so far. Most of them were discovered in the X-ray band, either from the emission of energetic bursts or pulsed emission with unique characteristics. Timing analyses of their persistent emission reveal that the magnetars spin slowly; their spin periods are in a tight range of $2-12$\,s and they experience large spin down rates ($3\times10^{-13}\sim 7.5\times10^{-10}$\,s\,s$^{-1}$). The dipole magnetic field strengths inferred from these spin properties are mostly of the order of $\sim 10^{14}$\,G \citep{ck98}. The luminosities of their persistent X-ray emission are between $10^{33}$$-$$10^{35}$\,erg\,s$^{-1}$, or lower. The emission of occasional energetic bursts has become a general characteristics of magnetars: bursts have been observed from all but three confirmed magnetars. The peak luminosities of magnetar bursts or flares vary in more than eight orders of magnitude ($10^{37}\sim>10^{45}$\,erg\,s$^{-1}$). Short bursts, which typically last for $\sim100$\,ms and with peak luminosities lower than $10^{41}$\,erg\,s$^{-1}$, are the most common population. 

Both persistent X-ray and occasional burst emission can be accounted for under the scheme of the magnetar model \citep{td95,td96}. In this picture, the magnetic field of the neutron star plays a crucial role in generating both types of emission. The persistent emission is expected from a hot spot on the neutron star surface heated by the decay of the strong magnetic field of the neutron star. The bursts might originate from the energy release either through fracturing of the neutron star crust due to large magnetic stress, or the split and reconnection of the oppositely oriented magnetic field lines \citep{lyutikov2003}. Magnetars occasionally undergo outburst episodes during which they experience remarkable increase in their persistent energy output, always in conjunction with the emission of energetic short bursts \citep{rea2011}.

In the magnetar model, small scale fracturing of the neutron star crust is invoked to explain a portion of the persistent X-ray emission, while large fractures can manifest themselves as bursts \citep{td96}. \citet{nakagawa2011} and \citet{enoto2012} studied the spectra in $1-300$\,keV of weak bursts from \sgr (31 events) and \sgrb (13 events), respectively, observed with \textit{Suzaku}. They investigated the stacked burst spectra with the sum of two blackbody and a power law functions and compared their spectral parameters to the results from the spectral fitting of persistent X-ray spectra with the phenomenological model (the sum of a blackbody and a power law function). Based on the similarity of the resulting power law indices (although with large errors), they suggested that persistent emission from magnetars can be due to a very large number of weak short bursts. We also previously studied spectral properties of the low fluence bursts and the persistent emission, both observed in a deep \textit{XMM-Newton} observation of \sgr on its most burst active day \citep{lin2012b}. Instead of phenomenological functions, we adopted physically motivated models for both persistent and burst spectra. In particular, we used the surface thermal emission with magnetosphere scattering model \citep[STEMS\footnote{The STEMS model assumes that the thermal emission from the surface of the neutron star is transfered through highly magnetized atmosphere before being scattered in the magnetosphere. The STEMS model has four parameters: magnetar surface temperature ($kT$) and magnetic field strength ($B$); the thermal electron speed ($\beta$) and the optical depth ($\tau$) of the scattering plasma in the magnetosphere.}, see][and references therein]{guver2007} and a modified blackbody \citep{lyubarsky2002} with resonant cyclotron scattering model \citep[MBB+RCS\footnote{The MBB+RCS model considers the burst spectrum as a thermal emission from a plasma bubble trapped in the magnetosphere which is modified by the strong magnetic field \citep{lyubarsky2002} and redistributed through the resonant cyclotron scattering by the warm plasma with the thermal electron speed $\beta$ and optical depth $\tau$ in the magnetosphere. Besides the scattering parameters, it has one more parameter $kT$, the temperature of the modified blackbody.},][]{lin2012b}. Both models include the same resonant cyclotron scattering by non-relativistic electrons in a warm magnetospheric plasma \citep{lyutikov2006}. Our spectral analyses showed that the persistent and burst emission are better described with the STEMS and MBB+RCS model, respectively, which indicated that the persistent X-ray emission and low fluence bursts originate through slightly different physical processes.   

It is, therefore, crucial to systematically study temporal and spectral properties of the low fluence magnetar bursts, in order to better understand the nature of these weak bursts and their connection with the persistent emission. However, identifying the very weak magnetar bursts using conventional techniques is a challenging task. A major fraction of these bursts are usually smoothed out in binned data. Also it is difficult to distinguish dim bursts from fluctuations of the persistent emission. As a result, the burst search methodology in our earlier study using binned data limited us to include even weaker bursts and understand their characteristics through detailed spectral and temporal analyses. 

A possible way to identify weak magnetar bursts is using the Bayesian blocks method. The Bayesian block representation of time series is a nonparametric method to detect local structures especially in highly variable data sets \citep{scargle98,scargle2013}. It can provide a simple description of the overall shape of a burst temporal profile as well as a direct measurement of the burst duration. This method has been adopted as a standard tool to calculate the duration of Gamma-ray Bursts (GRBs) observed with \textit{Swift}/BAT; also has been applied to search for the extended emission from short GRBs \citep{norris2010,norris2011,kaneko2013} and to calculate the duration of GRBs observed with \textit{Fermi}/GBM \citep{qin2013}. Moreover, the Bayesian blocks method has been used in the spatial domain to confirm the source detection in \textit{Fermi}/LAT data. 

In this paper, we introduce the Bayesian blocks method, for the first time, to identify very dim magnetar bursts using unbinned data of \sgr and \sgrb collected with \textit{XMM-Newton} and \textit{Swift}/X-ray telescope (XRT), respectively. We also investigate the temporal and spectral properties of the identified burst sample. This paper is formed in seven sections. In Section 2 and 3 we describe the data and our burst search algorithm, respectively. We investigate burst temporal properties in Section 4. In Section 5, we examine the accumulated spectra of dim bursts with physically motivated  emission models, and present the results of our detailed spectral analysis. We construct and evaluate the fluence distributions of the dimmest bursts for both sources in Section 6, and finally discuss the implications of our results in Section 7.

\section{The Sources and Observations}

In the past five years, only two magnetars, \sgr and \sgrbnos, experienced very active burst emitting episodes, during which tens or even hundreds of bursts had been observed with several X-ray and soft $\gamma$-ray telescopes. We investigated the observations of these two sources during their most burst active episodes. In our investigations, we focused on energies below 10\,keV, since the persistent emission was mainly detected in this energy range.

\subsection{\sgrnos}

\sgrnos, a magnetar located in the anti-Galactic center direction, was discovered with the detection of a series of short bursts starting on 2008 August 22 with \textit{Swift}, \textit{Fermi}/GBM, \textit{Konus-Wind}, and \textit{Suzaku} \citep{enoto2009, aptekar2009, kumar2010, nakagawa2011, lin2011}. The burst activity lasted for about two weeks. A spin period of $5.762$\,s and a spin-down rate of $\sim5.8\times10^{-12}$\,s\,s$^{-1}$ have been obtained from subsequent monitoring X-ray observations. These timing properties inferred a dipole magnetic field of $\sim2\times10^{14}$\,G \citep{rea2009, gogus2010}. 

The source has been observed seven times with \textit{XMM-Newton} \citep{jansen2001} within forty days following its discovery. The first of these pointed observations (Observation ID: 0560191501) spanned for 48.9 ks of 2008 August 23, the most burst active day of the source. This observation was carried out in small window mode of the European Photon Imaging Camera (EPIC) pn instrument \citep{struder2001}. The temporal resolution in this observing mode is 5.7 ms, that is fine enough to study short bursts. We selected the source counts from a circular region with a radius of $35\arcsec$ centered at the source location. The background region was selected with the same aperture size from a source free portion on the same chip. We processed the data using SAS version 11.0.0 with the latest calibration files generated on 2012 May 18. 

\subsection{\sgrbnos}

\sgrb is one of the few magnetars that have been detected in both radio and X-ray bands. The source was first identified in X-rays \citep{lamb1981}, and was suggested as an anomalous X-ray pulsar associated with a young supernova remnant G\,$327.24-0.13$ based upon its X-ray spectral properties \citep{gelfand2007}. Its magnetar nature has been confirmed with the detection of the radio pulsations with a period of 2.096\,s, the shortest one among all known magnetars, and a period derivative of $2.3\times 10^{-11}$\,s\,s$^{-1}$, which yields a dipolar magnetic field of $2.2\times10^{14}$\,G \citep{camilo2007}. One year later, its spin period was also revealed in the X-ray band with a deep \textit{XMM-Newton} observation \citep{halpern2008}. The burst activity from this source has been detected for the first time in 2008 October with both \textit{Swift}/BAT and \textit{Fermi}/GBM \citep{israel2010,vonkienlin2012}. On 2009 January 22, \sgrb became extremely burst active: hundreds of bursts were detected in one day with several high energy instruments \citep{mereghetti2009,savchenko2010,bernardini2011,dib2012,vdh2012}. The source returned back to quiescence after March 2009 \citep{vonkienlin2012}. During and after the burst active periods, both the persistent emission and timing properties of the source changed remarkably \citep{enoto2010,ng2011,bernardini2011,scholz2011,dib2012,kuiper2012}. It has also been reported that \sgrb is embedded in a dust scattering halo \citep{vink2009,tiengo2010,olausen2011}. 

Following its identification as a magnetar, \sgrb has been monitored with \textit{Swift}/XRT. In this work, we selected two observations (00340573000 \& 00340573001), both performed on 2009 January 22\footnote{The observation 00340573000 started at 02:26:22 and stopped at 07:30:19. The second observation 00340573001 was performed from 09:18:28 to 17:18:09. All times are in UT.} with a total exposure time of 15.8 ks. These observations were carried out in window timing mode, which sacrifices one dimension of the spatial image for the high temporal resolution of 1.766 ms. We selected the source region as a 34 pixel slide centered at the source position which corresponded to $80\arcsec$. Due to the one dimension image and the wide PSF, the event list from the selected source region is actually the combination of the emission from the source, dust halo and sky background. In order to eliminate the contribution of the dust halo and the sky background on the burst spectra, we extracted the background spectrum from the same region during the burst-free time intervals.

\section{Identification of short bursts with Bayesian blocks representation}

In the Bayesian blocks method, data are represented with a series of step functions, which can be in different sizes, determined by maximizing the block likelihood function. For the photon counting instruments, such as \textit{Swift}/XRT and \textit{XMM-Newton}, the detection probability of one event follows Poisson statistics. The likelihood function ($L$) for block $k$ is 
\begin{equation}
ln L^{(k)} = N^{(k)} ln \lambda^{(k)} - \lambda^{(k)} T^{(k)}
\end{equation}
where $N^{(k)}$ is the total number of events in block $k$, $\lambda^{(k)}$ is the expected count rate, and $T^{(k)}$ is the duration of block $k$. 

A priori information in this method is the improvement of the likelihood function by adding one more block. It depends on the number of events, as well as the false positive detection rate \citep[see Equation 21 in][]{scargle2013}. \citet{scargle2013} calibrated this value with large sets of simulations, and also showed that the detection sensitivity of this method can approach the ideal theoretical limits. The Bayesian blocks method has three advantages in identifying short magnetar bursts. First, it avoids placing a priori limitation on the scale and resolution of the search. It is sensitive to short weak bursts which may be smoothed out when the data are binned. Second, it is applicable to data with wide ranges of amplitude, time scales and noise levels. Finally, it is possible to determine the duration of short bursts independent of any background modeling.

We adopted the algorithm provided in \citet{scargle2013}, and generated automatic procedures to search for the bursts using the list of time-tagged events in a specified energy range, i.e., $0.2-10$\,keV for the \textit{XMM-Newton} data and $0.3-10$\,keV for \textit{Swift}/XRT data. Our burst search procedure included two rounds of Bayesian block representation. The first round was a blind search through the whole data set and aimed to find candidate burst blocks. The second round of Bayesian block representation was performed to confirm the candidate burst blocks, along with sufficient burst-free intervals. 

To optimize the computation time for the first round search, we employed a moving box-car approach: we calculated the Bayesian block representation for a 20 s long observing window, starting from the beginning of the observation, and shifted forward by 10 s in each time. This way, the entire exposure span, except for the first and last 10 s segments of each continuous time interval, was covered twice by consecutive observing windows in our first round of search. We then selected all blocks with durations less than the spin period of the source as candidate burst blocks. We excluded the candidate blocks which lined up with a boundary of any 20 s segment, as they will be picked in the middle of the previous or the next segment.
 
Before the second round representation, we further gathered the candidate burst blocks into groups, in which the separation between any two adjacent blocks in the time series is smaller than 10 s. Then, we performed the second round Bayesian blocks representation on each group, using the time-tagged events within the time segment from 10 s before the first block to 10 s after the last block. In this way, we can confirm the candidate burst blocks with sufficient underlying emission (background) in both pre- and post-burst intervals and remove the duplicates. Additionally, the policy we followed to form the groups guaranteed that all underlying emission time intervals are burst free. 

We firmly identified the magnetar bursts after two rounds of Bayesian blocks representation as follows:
We first selected the blocks with duration longer than 6 s as background and calculated the background count rate using the weighted mean of the background block rates. We then identified the blocks with duration less then the spin period of the source and count rate higher than the background level as burst blocks. Finally, we combined any continuous burst blocks as parts of a single magnetar burst.

\subsection{\sgrnos}

By applying the Bayesian blocks algorithm to the $0.2-10$\,keV data of \textit{XMM-Newton} observation of \sgrnos, we found 320 bursts. Due to the possible pile-up effects, we excluded 58 events that coincide with any one second segment during which more than 50 counts were accumulated. Therefore, we have 262 bursts\footnote{There are two bursts in this sample having high block count rates, over 600\,counts\,s$^{-1}$, and they may have suffered by $\sim5\%$ pile-up. We kept these two bursts in the sample since they would not significantly affect the accumulative analyses results. } for further analyses. 

We studied the low-fluence burst stacked spectrum using the same observation in \citet{lin2012b}, where we excluded all the 100 ms bins with no more than 10 counts from the burst spectrum, due to the fact that it was difficult to distinguish the burst emission from the fluctuations of the persistent source in this flux regime. Comparing these two samples, all intervals considered as bursts in \citet{lin2012b} are identified as bursts with the Bayesian blocks algorithm. Moreover, at least $60\%$ (165/262) of bursts identified with the Bayesian blocks method are too dim to be determined as bursts in the previous study, and are analyzed for the first time here. These results show that our search method is successful in detecting of both bright and dim bursts, and also capable of identifying even weaker bursts from the fluctuations of the persistent emission. 

\subsection{\sgrbnos}

We searched for bursts from \sgrb in two \textit{Swift}/XRT observations in the $0.3-10$\,keV energy range using the same Bayesian blocks method. In the first observation (ID: 0034057300), we found more than 150 burst blocks grouped in one segment lasting $\sim230$ s for the second round of the Bayesian block representation. The duration of this segment is $\sim10$ times longer than most other segments for the second round search for both sources. In order to avoid a much higher threshold due to the large number of counts, we forced a priori information to remain at the initial value for this segment. We identified a total of 301 bursts in this observation and 103 bursts in the following one (ID: 0034057301). The data collected in window timing mode are not affected by photon pile-up for count rates less than 100\,counts\,s$^{-1}$ \citep{romano2006}. We eliminated the bursts that exceeded this limit. As a result, we have had 193  and 82 bursts in the observations 0034057300 and 0034057301, respectively, for further studies.

\section{Temporal Properties}

\subsection{Bayesian blocks profile}

\begin{figure}[h]
    \includegraphics[scale=0.9]{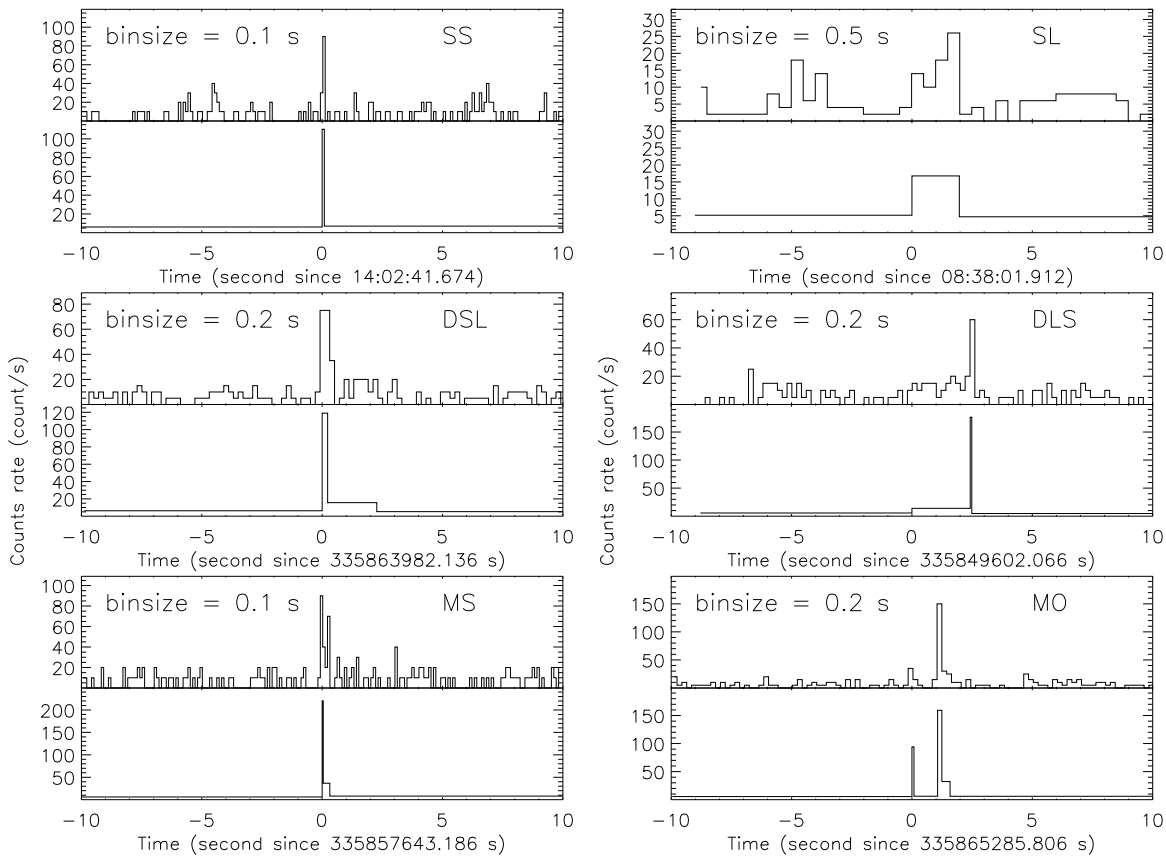}
\caption{Examples of six common types of Bayesian blocks profile of bursts from \sgrnos. For each example, the upper panel shows the lightcurve of the burst and its corresponding Bayesian blocks profile is shown in the lower panel.   \label{fig-lc}}
\end{figure}

The Bayesian blocks representations of the temporal profiles of magnetar bursts are much simpler compared with their lightcurves. Figure \ref{fig-lc} shows lightcurves and corresponding Bayesian blocks profiles of bursts from \sgrnos, representing six common types that emerge from our analysis. The burst lightcurves are binned with different binsizes for clarity. The majority of magnetar bursts can be represented with only one block. The burst block is either a short peak less than half a second (SS: single short) or otherwise a broad bump over the local level of the persistent emission (SL: single long). Some bursts have at least two subsequent blocks in their profiles and all of them are short ones (i.e., $< 0.5$\,s, MS: multiple short). The rest of bursts have at least one long block ($> 0.5$\,s). We divided them into three groups: DSL (double short+long) bursts are two blocks events which have a single peak with extended emission; DLS (double long+short) bursts also have two blocks, but with an enhanced emission leading to the single peak; and MO (multiple others) events include all other cases. We summarized the basic morphological properties, as well as the distribution of burst patterns among the two sources in Table \ref{tab-bbtype}.

\begin{deluxetable}{lccccccc}
\tabletypesize{\scriptsize}
\setlength{\tabcolsep}{0.1in} 
\tablecaption{The summary of six types of Bayesian blocks profiles of magnetar bursts. \label{tab-bbtype}}
\tablewidth{0pt}
\tablehead{\colhead{Morphology} & \colhead{SS} & \colhead{SL} & \colhead{DSL} & \colhead{DLS} & \colhead{MS} & \colhead{MO}&\colhead{N\tablenotemark{a}}}
\startdata
\sgr \textit{XMM-Newton} & $201 (76.7\%)$ & $40 (15.3\%)$ & $12 (4.6\%)$ & $2 (0.8\%)$& $6 (2.3\%)$ & $1 (0.4\%)$ & $320$\\
\sgrb 00340573000 orbit1 & $32 (71.1\%)$ & $8(17.8\%)$ & $1(2.2\%)$ & 0 & $2(4.4\%)$ & $2(4.4\%)$ & $59$\\
\sgrb 00340573000 orbit2 & $50(47.6\%)$ & $18(17.1\%)$ & $9(8.6\%)$ & $1(1\%)$ & $16(15.2\%)$ & $11(10.5\%)$ & $152$\\
\sgrb 00340573000 orbit3 & $31(75.6\%)$ & $7(17.1\%)$ & 0 & 0 & $3(7.3\%)$ & 0 & $62$\\ 
\sgrb 00340573000 orbit4 & $2(100\%)$ & 0 & 0 & 0 & 0 & 0 & $28$\\ 
\sgrb 00340573001 & $65(79.3\%)$ & $6(7.3\%)$ & $5(6.1\%)$ & 0 & $5(6.1\%)$ & $1(1.2\%)$ & $103$\\ 
\sgrb all & $180(65.5\%)$ & $39(14.2\%)$& $15(5.5\%)$ & $1(0.4\%)$ & $26(9.5\%)$ & $14(5.1\%)$ & 404 \\
\enddata
\tablenotetext{a}{N: Total number of burst (including piled-up bursts)}

\end{deluxetable}

\subsection{Duration}

In the earlier investigations of the temporal properties of magnetar bursts, burst durations were estimated by fitting the cumulative count or energy fluence lightcurves with the sum of a step function and a first order polynomial. In this formalism, the step function provides the amplitude of the burst, and the polynomial represents the behavior of the background emission \citep[e.g.,][]{gogus2001,lin2011}. This method, however, leaves room for uncertainties due to the human intervention. In particular, the choice of background intervals before and after the burst might be subjective and, in turn, might result with slightly different duration estimation. In this study, we define the duration of a burst with the time length of representing Bayesian blocks: the duration of a burst is the interval from the start of the first burst block to the end of the last one. Since the change point (boundary) between the background and burst blocks as settled with a certain threshold \citep{scargle2013}, the duration calculation based on the Bayesian blocks profile is simple, direct and not influenced by any subjective background selections. Note here the fact that due to difficulties arising from background fluctuations, $T_{90}$ or $T_{50}$ durations have been conventionally calculated for bursts \citep{ck1993}. The Bayesian burst blocks, however, represent the entire burst durations.
  
\subsubsection{\sgrnos}

\begin{figure}[h]
    \includegraphics[scale=0.4,bb=10 230 550 630]{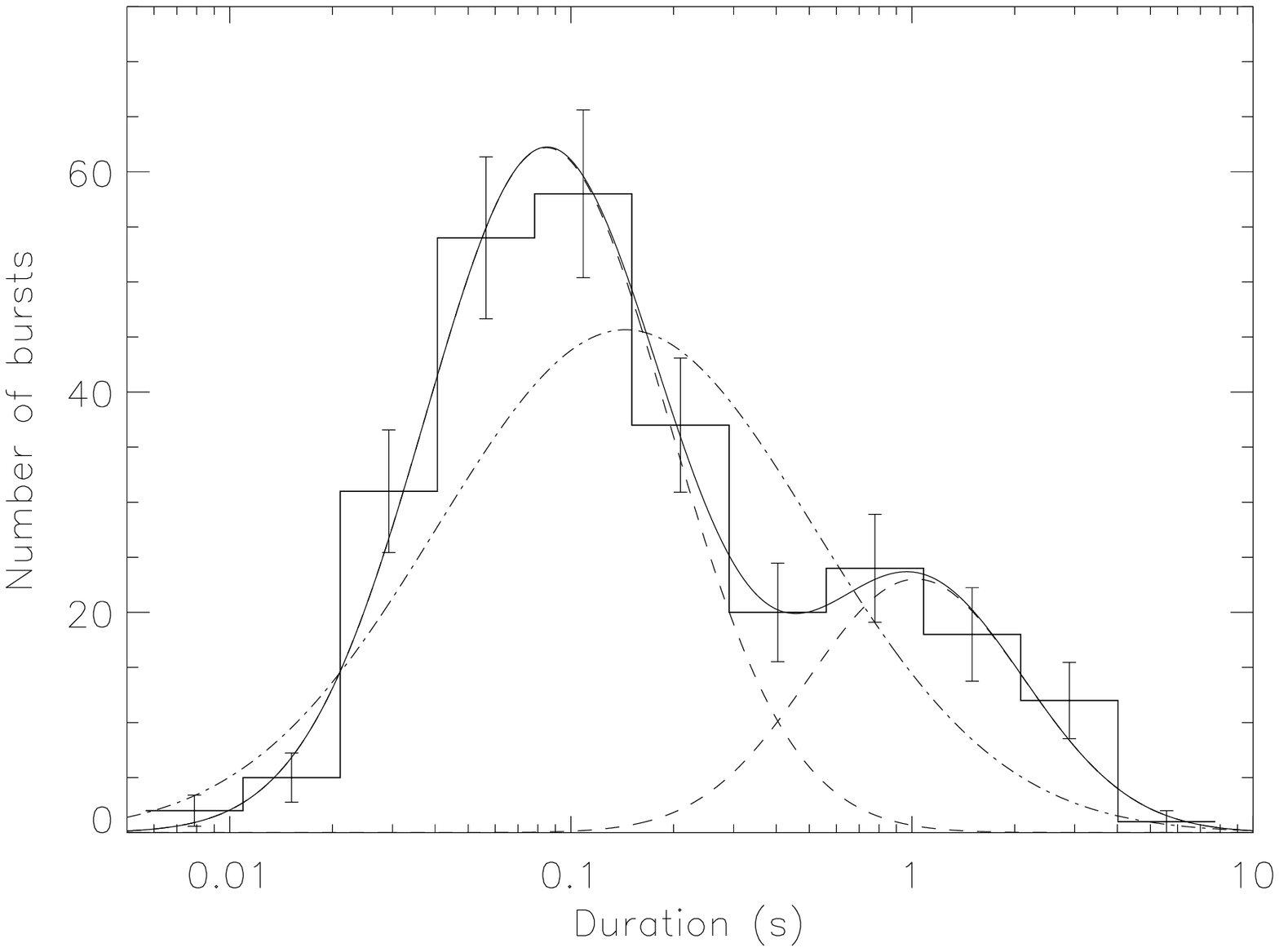}
    \includegraphics[scale=0.4,bb=10 230 550 630]{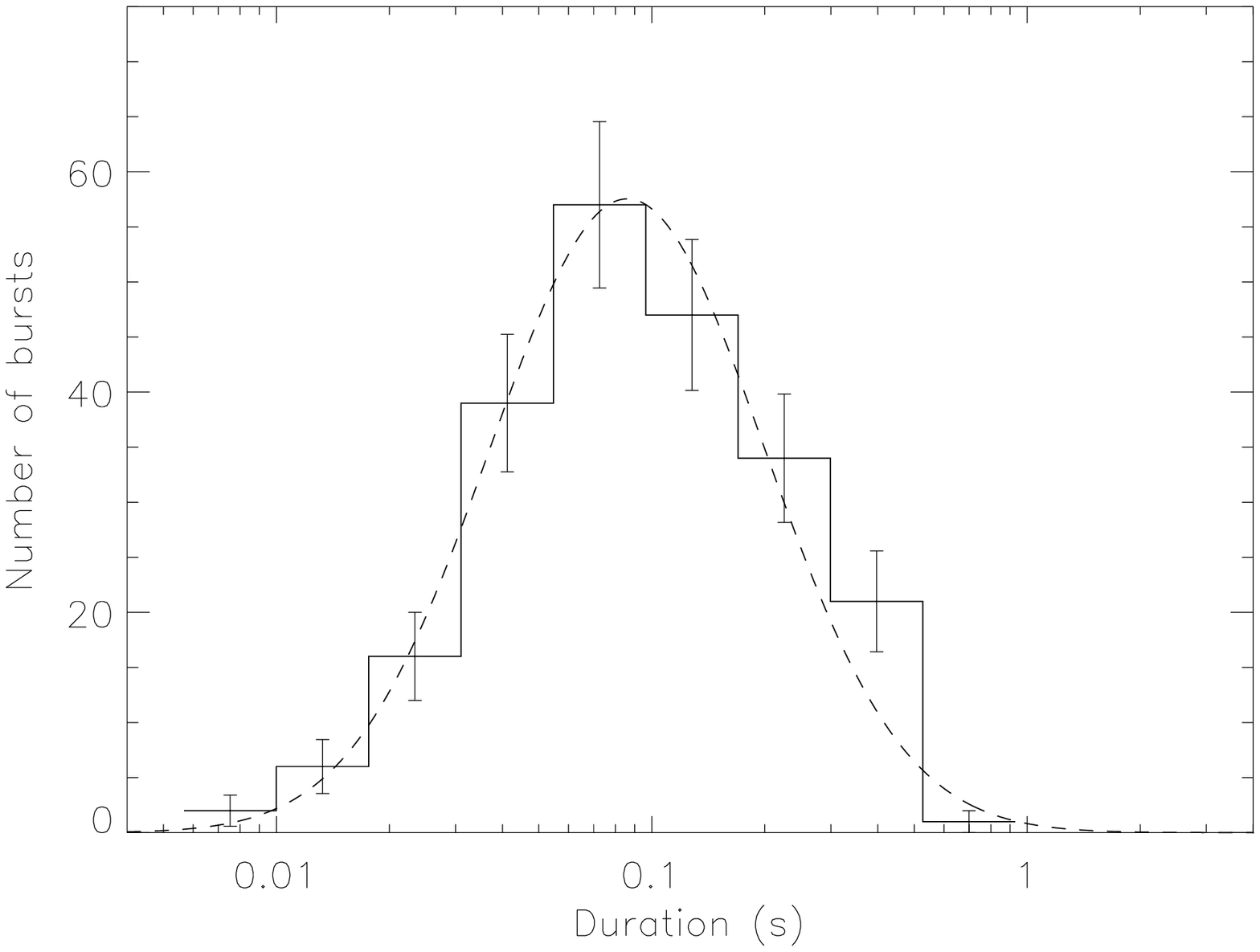}
\caption{\textit{Left:} the distribution of duration of 262 non piled-up bursts of \sgrnos. The solid curve presents the best fit of the histogram with a sum of two Gaussian functions. Two components are also shown with dashed curves. The dash-dotted curve exhibits the best fit single Gaussian function to the distribution. \textit{Right:} The duration distribution of short burst blocks (SS \& MS, and short blocks of DSL, DLS, MO bursts) of \sgrnos. The best fit Gaussian function is presented with a dashed curve. \label{fig-durdis}}
\end{figure}

We present the duration distribution of all 262 non piled-up \sgr bursts in the left panel of Figure \ref{fig-durdis}. The apparent deviation from a log$-$Gaussian shape reflects also into the quantitative case: a fit to the duration distribution of \sgr with a Gaussian function yields a $\chi^2$ of 41.14 for 8 degrees of freedom (dof). We then fit the distribution with the sum of two Gaussian functions, which resulted in a very good fit ($\chi^2/dof = 3.65/5$). We estimate the chance probability of the preference of the more complicated model as $1\times10^{-8}$. The best fit mean values of the two Gaussian components are $85\pm8$\,ms with $\sigma=0.36\pm0.03$ and $1028^{+220}_{-181}$\,ms with $\sigma=0.32\pm0.05$ (both $\sigma$ are in decades). The distribution reaches a local minimum value at 456\,ms in between the two peaks.

The second Gaussian function is dominated by the long blocks, namely SL bursts and long block of DSL, DLS, MO events. We excluded those bursts/blocks, and generated the distribution of only short bursts/blocks shown as the right panel of Figure \ref{fig-durdis}. This distribution is fit well with a single Gaussian function ( $\chi^2/dof=9.15/6$), and the best fit mean value is $86\pm5$\,ms, with $\sigma=0.37\pm0.02$. As expected, this set of parameter agrees well with that of the Gaussian component with the smaller mean value.

\subsubsection{\sgrbnos}
\begin{figure}[h]
\includegraphics[scale=0.4,bb=10 230 550 630]{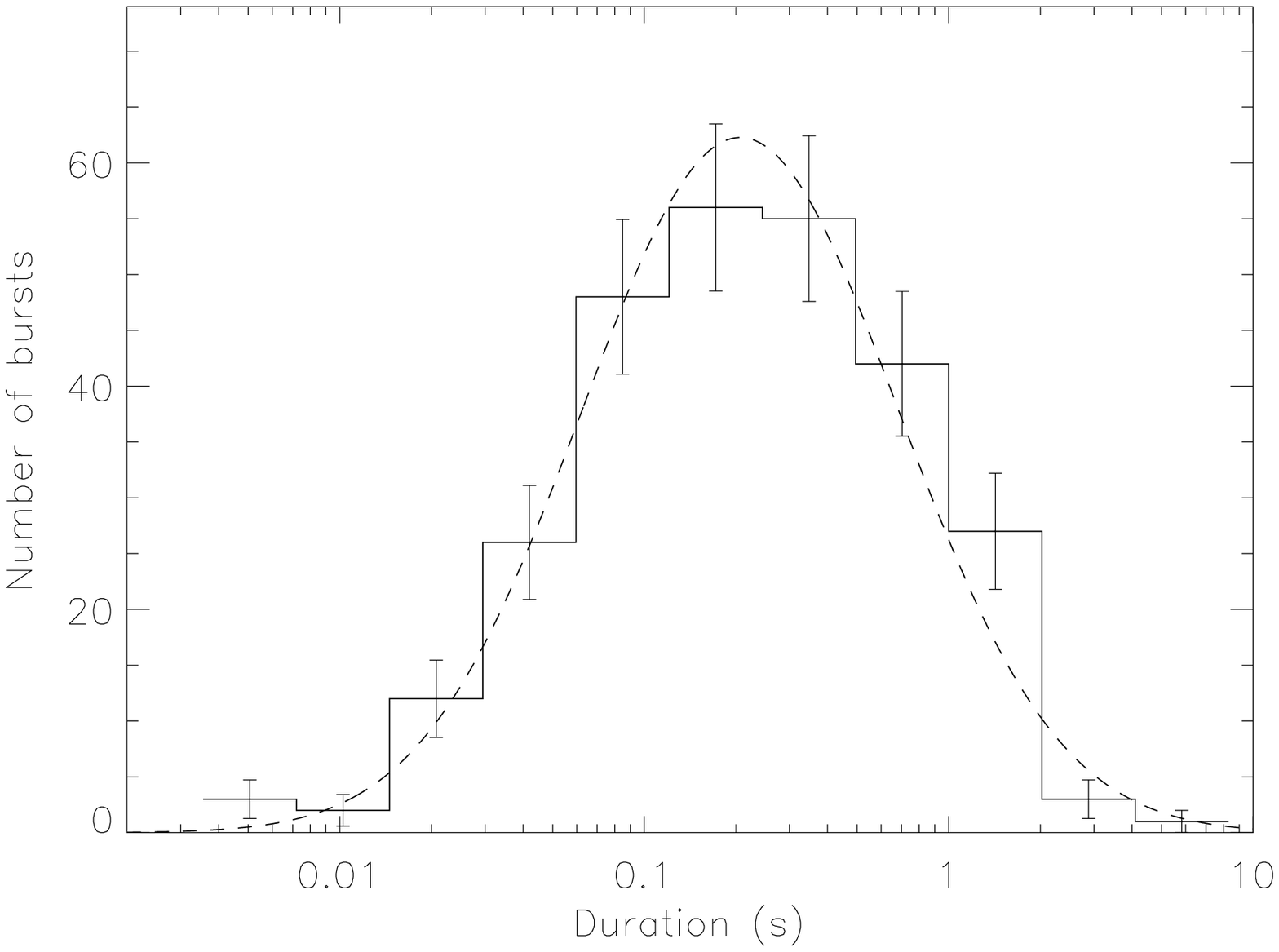}
\includegraphics[scale=0.4,bb=10 230 550 630]{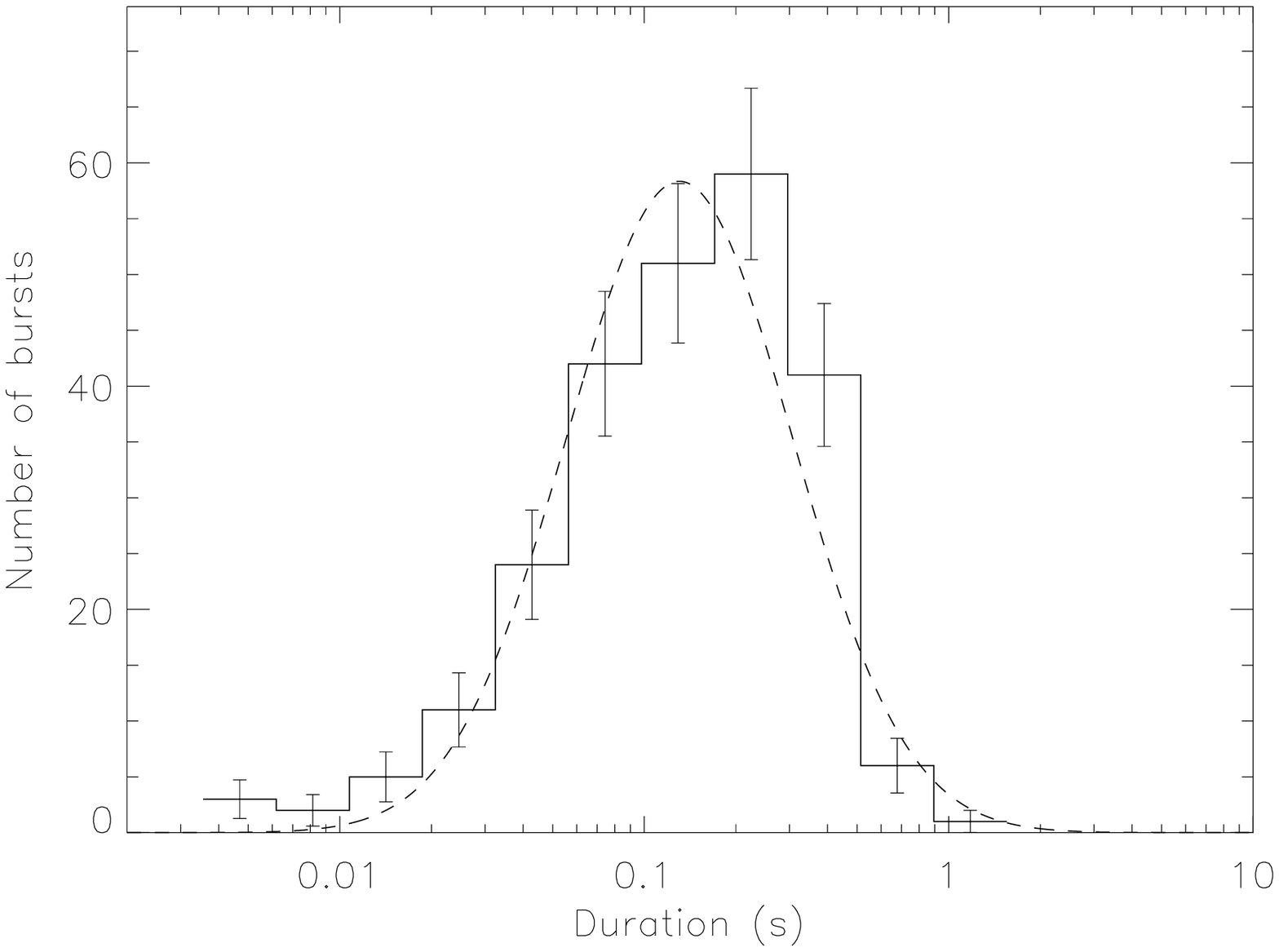}
\caption{Duration distribution of 275 non piled-up bursts from \sgrb (\textit{left}) and 245 short bursts/blocks (SS \& MS bursts, and short blocks of DSL, DLS, MO events, \textit{right}). The dashed curve in each panel exhibits the best fit Gaussian function to the histogram. \label{fig-durdisb}}
\end{figure}

We studied the duration distribution of \sgrb using 275 bursts. The distribution (the left panel of Figure \ref{fig-durdisb}) can be fit well with a Gaussian function in the logarithmic space ( $\chi^2/dof=9.83/8$) with a mean value of $207^{+16}_{-15}$\,ms and $\sigma=0.52\pm0.02$. Unlike the distribution of \sgrnos, a second Gaussian function is not required here. We also generated the duration distribution of 245 short bursts/blocks including SS, MS and short blocks of DSL, DLS, MO bursts (23 short bursts are contributed by MO events). This distribution can also be described with a Gaussian function as shown in the right panel of Figure \ref{fig-durdisb}. The best fit mean value is $131\pm8$\,ms with $\sigma=0.37\pm0.02$ ($\chi^2/dof=19.10/8$). 

\section{Spectral properties}

\subsection{\sgrnos}

In \citet{lin2012b}, we studied the stacked spectrum of 129 relatively dim \sgr bursts. With a larger sample from the Bayesian blocks burst search, we are now able to look into the spectral behavior of the weaker bursts, grouped in morphological type and fluence. For this purpose, we calculated a signal to noise ratio (SNR) for each block as $(N_T-N_B)/\sqrt{N_T}$, where $N_T$ and $N_B$ are the total and background counts collected within the block, respectively. This value is a general indicator of the burst fluence, and ranges in between 1.6 and 6.3. Then, we further divided SS bursts into three groups with SNR $> 4$, $3<$SNR$\le4$, and SNR $\le3$. The SL bursts were set into two groups with the SNR boundary of 4. For DSL \& DLS events, we merged all their short blocks into one spectral group and the long blocks into another one. We kept MS bursts in one group and excluded the MO event since there was only one such burst. Overall, we have formed eight spectral groups of bursts with sufficient number of counts, different Bayesian block profiles and energetics, and extracted a stacked spectrum for each group. 

\begin{figure}[h]
\includegraphics[scale=0.8,bb=30 150 580 500]{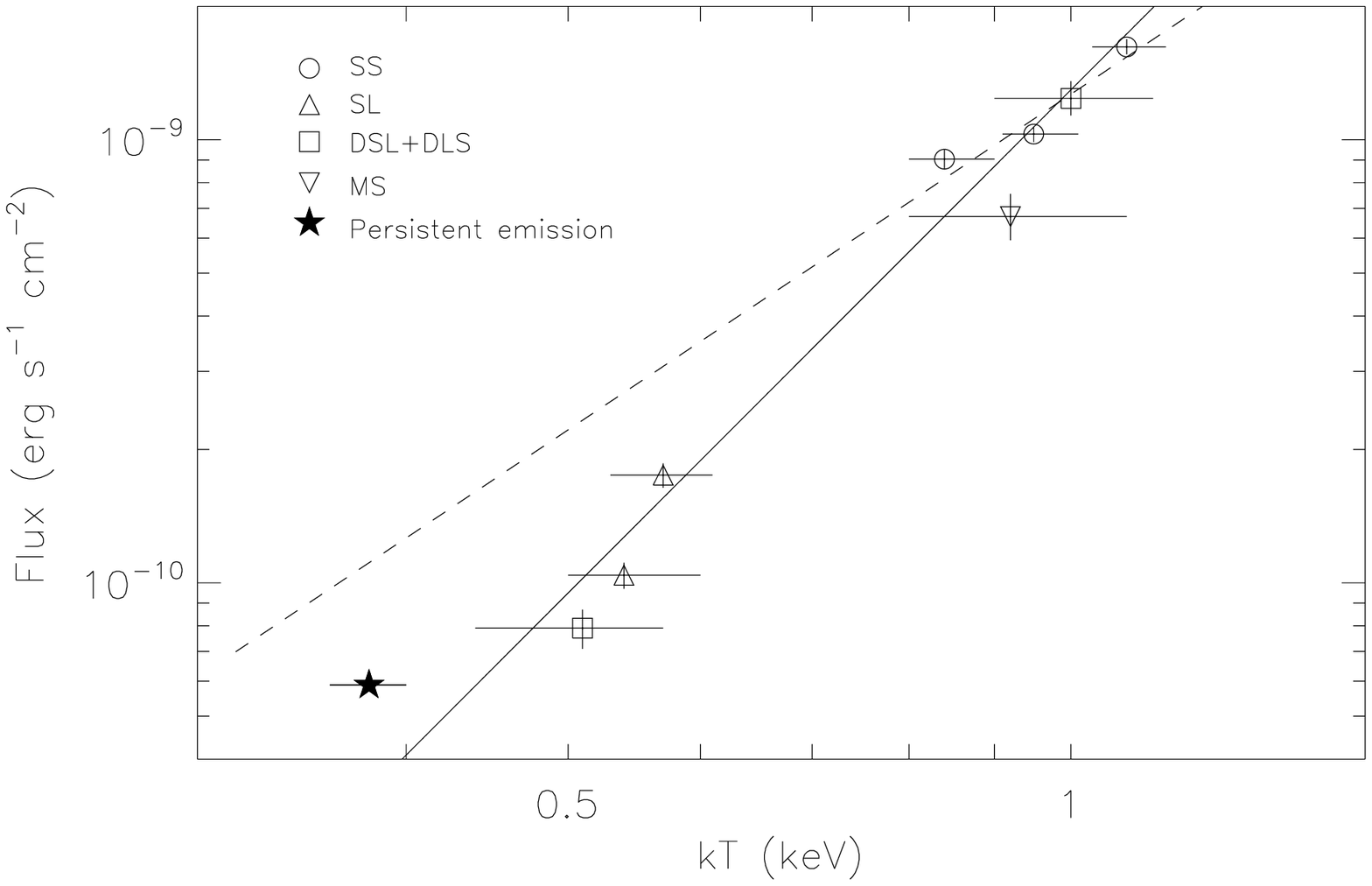}
\caption{The correlative trend between the unabsorbed flux in $0.5-10$\,keV and the temperature of modified blackbody for \sgr bursts. The spectral results of SS, SL, DSL+DLS, MS bursts are shown as circles, triangles, squares, and upside down triangles, respectively. The five-point star is the STEMS fit result of the persistent emission spectrum taken from \citet{lin2012b}. The solid line is the best fit power law trend using all burst spectra, and the dashed line exhibits the best fit power law using only the spectra of SS \& MS bursts and short blocks of DSL \& DLS bursts.    \label{fig-ktflux}}
\end{figure}

We fit each stacked spectrum with the MBB+RCS model. We set the interstellar absorption and magnetospheric scattering parameters of the plasma at the values obtained from the persistent emission spectral analysis ($n_{\rm H} = 6.7\times 10^{21}$\,cm$^{-2}$, $\beta = 0.37$, $\tau = 5$, Lin et al. 2012). We found that all stacked burst spectra can be fit well with the MBB+RCS model, for which we listed the best fit model parameters in Table \ref{tab-bstspec}. We noticed that the unabsorbed flux in $0.5-10$\,keV is correlated with the temperature of the modified blackbody (see Figure \ref{fig-ktflux}). The Spearman's rank-order correlation coefficient is $0.98$ with a random occurrence probability of $3.3\times10^{-5}$. The correlative trend between the flux and temperature can be expressed as a power law with $Flux\propto kT^{3.8\pm0.3}$ ($\chi^2/dof=4.2/6$; as shown with the solid line in Figure \ref{fig-ktflux}). The power law index becomes flatter ($Flux \propto kT^{2.5\pm1.0}$ with $\chi^2/dof=1.3/3$) when we only include spectral results of SS \& MS bursts and short blocks of DSL \& DLS events. 

We also fit the stacked spectra with STEMS model, which is preferred by the persistent emission spectral shape. We found that the stacked spectra of SL bursts and long blocks of DSL \& DLS bursts can also be fit well with the STEMS model, while other types cannot be fit with it. We listed the STEMS model parameters in Table \ref{tab-bstspec}. The temperatures of long block bursts are consistent with one another ($\sim$0.4 keV) within errors.  Additionally, we showed the temperature of the hot spot on the neutron star surface as obtained by fitting its persistent emission from the same observation \citep{lin2012b} with a five-point star in Figure \ref{fig-ktflux}.

\begin{deluxetable}{cccccc}
\tabletypesize{\scriptsize}
\setlength{\tabcolsep}{0.1in} 
\tablecaption{Spectral fit results. \label{tab-bstspec}}
\tablewidth{0pt}
\tablehead{\colhead{Type-(SNR)} & \colhead{Exposure} & \colhead{kT} & \colhead{Flux\tablenotemark{a}} &  
\colhead{$\chi^2/d.o.f$} &  \colhead{Burst energy per count} \\
\colhead{} & \colhead{(s)} & \colhead{(keV)} & \colhead{($10^{-10}$\,erg\,s$^{-1}$\,cm$^{-2}$)} & \colhead{}  & \colhead{($10^{-12}$\,erg\,count$^{-1}$)}
} 
\startdata
\multicolumn{6}{c}{\sgrnos}\\
\hline
\multicolumn{6}{c}{ }\\
\multicolumn{6}{l}{MBB+RCS, $n_{\rm H}=6.7\times10^{21}$\,cm$^{-2}$, $\beta=0.37$, $\tau=5$}\\
SS$-(\le3)$ & 4.556 & $0.84_{-0.04}^{+0.06}$ & $9.0\pm0.5$ & $44.64/37$ & $7.4$ \\
SS$-(3\sim4)$ & 6.391 & $0.95_{-0.04}^{+0.06}$ & $10.3\pm0.4$ & $61.93/51$ & $7.7$ \\
SS$-(>4)$ & 4.949 & $1.08_{-0.05}^{+0.06}$ & $16.2^{+0.7}_{-0.6}$ & $44.17/57$ & $8.3$ \\
SL$-(\le4)$ & 22.91 & $0.54_{-0.04}^{+0.06}$ & $1.0\pm0.1$ & $34.10/34$ & $6.6$ \\
SL$-(>4)$ & 15.33 & $0.57\pm0.04$ & $1.8\pm0.1$ & $24.16/31$ & $6.9$ \\
DSL, DLS$-short$ & 1.399 & $1.00^{+0.12}_{-0.10}$ & $12.4^{+1.2}_{-1.1}$ & $4.29/12$ & $8.7$ \\
DSL, DLS$-long$ & 18.35 & $0.51^{+0.06}_{-0.07}$ & $0.8\pm0.1$ & $22.27/21$ & $6.9$\\
MS & 1.662 & $0.92^{+0.16}_{-0.12}$ & $6.7\pm0.8$ & $4.13/7$ & $9.1$\\
\multicolumn{6}{l}{STEMS, $n_{\rm H}=6.7\times10^{21}$\,cm$^{-2}$, $\beta=0.37$, $\tau=5$, $B=2.2\times10^{14}$\,G}\\
SL$-(\le4)$ & 22.91 & $0.35_{-0.03}^{+0.07}$ & $1.1\pm0.1$ & $30.71/34$ & \nodata \\
SL$-(>4)$ & 15.33 & $0.42_{-0.02}^{+0.03}$ & $1.8\pm0.1$ & $29.11/31$ & \nodata \\
DSL, DLS$-long$ & 18.35 & $0.40^{+0.02}_{-0.08}$ & $0.8\pm0.1$ & $23.31/21$ & \nodata \\
\hline
\multicolumn{6}{c}{\sgrb XRT observation ID 00340573000} \\
\hline
\multicolumn{6}{c}{ }\\
\multicolumn{6}{l}{MBB+RCS, $n_{\rm H}=3.4\times10^{22}$\,cm$^{-2}$, $\beta=0.2$, $\tau=3$} \\
orbit1, SS & 12.71 & $2.64^{+0.22}_{-0.19}$ & $132.4^{+5.3}_{-5.2}$ & $80.71/75$ & $142.2$ \\
orbit2, SS &8.121 & $3.93^{+0.61}_{-0.45}$ & $224.2^{+10.3}_{-10.0}$ & $85.72/72$ & $160.1$ \\
orbit3, SS & 9.382 & $2.08^{+0.18}_{-0.15}$ & $112.5^{+5.4}_{-5.2}$ & $44.43/60$ & $133.6$ \\
orbit1, SL & 6.254 & $2.24^{+0.38}_{-0.28}$ & $67.0^{+5.5}_{-5.2}$ & $16.88/19$ & $147.3$ \\
orbit2, SL & 15.77 & $2.26^{+0.21}_{-0.18}$ & $58.6^{+3.0}_{-2.9}$ & $46.01/47$ & $136.2$ \\
orbit3, SL & 5.985 & $1.52^{+0.18}_{-0.16}$ & $62.3\pm4.6$ & $19.51/26$ & $122.9$ \\
orbit2,DSL, DLS short & 1.456 & $3.38^{+0.76}_{-0.51}$ & $370.3^{+29.3}_{-28.1}$ & $11.18/20$ & $165.3$ \\
orbit2,DSL, DLS long & 10.04 & $2.12^{+0.50}_{-0.33}$ & $32.2^{+3.6}_{-3.2}$ & $21.65/16$ & $143.9$ \\
orbit2, MS & 5.558 & $5.49^{+2.02}_{-1.04}$ & $286.4^{+16.0}_{-15.4}$ & $41.36/58$ & $176.0$ \\
orbit3, MS & 1.002 & $2.65^{+1.03}_{-0.55}$ & $218.3^{+29.9}_{-26.4}$ & $11.80/8$ & $160.4$ \\
orbit2, MO short & 4.463 & $3.47^{+0.53}_{-0.39}$ & $256.2^{+13.8}_{-13.3}$ & $41.23/47$ & $157.7$ \\
orbit2, MO long & 16.74 & $1.80^{+0.33}_{-0.26}$ & $16.7^{+1.7}_{-1.6}$ & $13.56/18$ & $132.7$ \\
\hline
\multicolumn{6}{c}{\sgrb XRT observation ID 00340573001} \\
\hline
\multicolumn{6}{c}{ }\\

\multicolumn{6}{l}{MBB+RCS, $n_{\rm H}=2.8\times10^{22}$\,cm$^{-2}$, $\beta=0.2$, $\tau=3$} \\
SS & 8.916 & $2.66^{+0.20}_{-0.17}$ & $179.8^{+6.9}_{-6.8}$ & $59.39/78$ & $135.7$ \\
SL & 4.565 & $2.05^{+0.67}_{-0.39}$ & $43.7^{+6.3}_{-5.4}$ & $8.38/11$ & $132.7$ \\
DSL long & 5.438 & $1.52^{+0.35}_{-0.24}$ & $30.5^{+3.7}_{-3.5}$ & $16.08/11$ & $115.0$ \\
MS & 2.234 & $3.60^{+2.62}_{-0.93}$ & $142.3^{+23.1}_{-18.8}$ & $8.32/12$ & $168.3$ \\

\enddata
\tablenotetext{a}{Average unabsorbed flux in the $0.5-10$\,keV energy range.}

\end{deluxetable}

\subsection{\sgrbnos}

The persistent source properties of \sgrb changed significantly during its burst active episodes. The source flux (count rate) in observation 00340573000 has varied by a factor of $\sim$10 from the first to the last orbit. Therefore, instead of averaging the whole observation, we extracted the burst and persistent emission spectra for each orbit separately for 00340573000. Since the source flux did not show any dramatic change in the later observation and the number of bursts is relatively small, we extracted spectra from the entire observation of 00340573001. We investigated burst spectra of \sgrb in morphological groups only since there are not enough number of bursts in each orbit to allow grouping in fluence. We listed all stacked spectra extracted and analyzed in Table \ref{tab-bstspec}.

Because of the existence of the dust scattering halo \citep{tiengo2010,vink2009,ng2011}, the choice of the interstellar absorption value for this source was not trivial. Therefore, we first performed simultaneous fit to the stacked spectra of all types of bursts in both observations with the MBB+RCS model and forced them to have the same common $n_{\rm H}$. \citet{ng2011} studied the persistent emission spectra collected before, during and after the burst active periods using the STEMS model. Considering the fact that the magnetospheric properties would not change over the short time scales of bursts, we fixed the scattering parameters in the MBB+RCS model at their resulting values for the burst active episode: the average speed of the electrons $\beta\sim0.2$ and the optical depth $\tau\sim3.0$ \citep{ng2011}. We obtained the best fit $n_{\rm H}$ as $(3.4\pm0.2)\times10^{22}$\,cm$^{-2}$ ($\chi^2/dof=434.67/466$) and $(2.8\pm0.3)\times10^{22}$\,cm$^{-2}$ ($\chi^2/dof=92.16/111$) for observation 00340573000 and 00340573001, respectively. Note that there are three different $n_{\rm H}$ values for \sgrb reported in the literatures using fine spatial resolution \textit{XMM-Newton} and \textit{Chandra} observations:  $(3.1\pm0.3)\times10^{22}$\,cm$^{-2}$ \citep{halpern2008}, $(2.8\pm0.3)\times10^{22}$\,cm$^{-2}$ \citep{vink2009} and $(4.5\pm0.1)\times10^{22}$\,cm$^{-2}$ \citep{ng2011}. Our results fall within the range of reported $n_{\rm H}$ estimates. Therefore, we performed our detailed spectral analysis with the MBB+RCS model for each stacked spectrum using galactic absorption value of $3.4\times10^{22}$\,cm$^{-2}$ for observation 00340573000 and $2.8\times10^{22}$\,cm$^{-2}$ for the other one. The best fit parameters as well as the fit statistics are presented in Table \ref{tab-bstspec}. The temperature of the modified blackbody spreads between 1.5\,keV and 5.7\,keV, much higher than those of \sgrnos, and the average unabsorbed flux in $0.5-10$\,keV of bursts is also much larger. 

\begin{figure}[h]
\includegraphics[scale=0.8,bb=30 150 580 500]{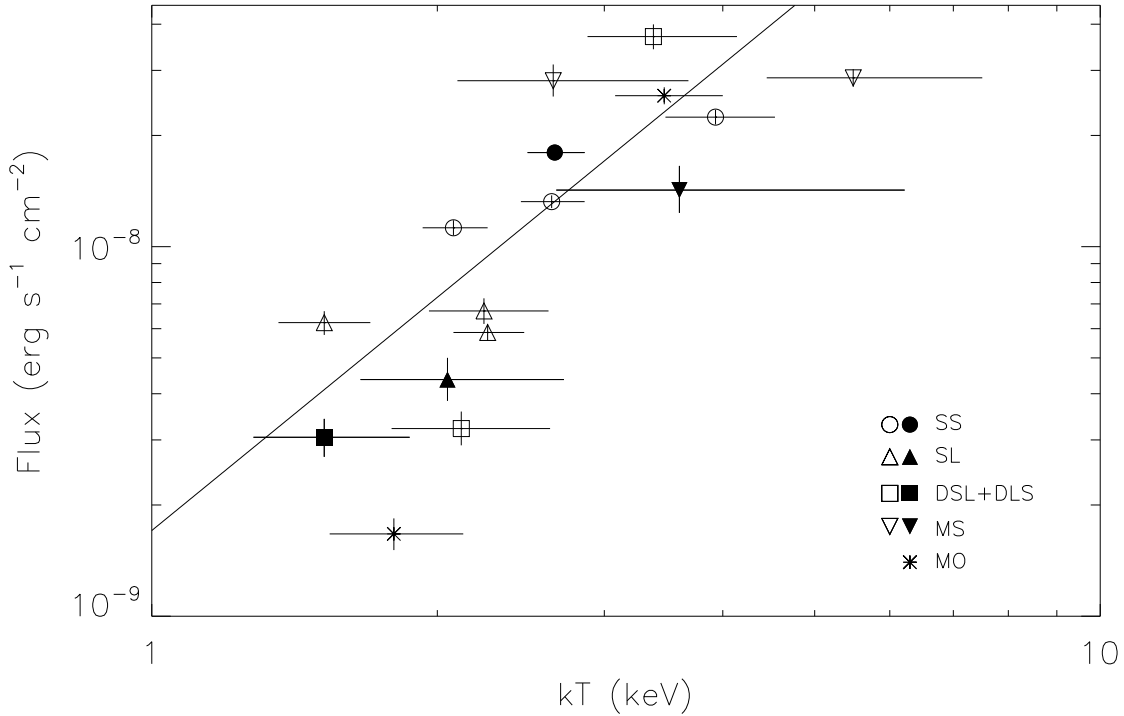}
\caption{The correlative trend between the unabsorbed flux in $0.5-10$\,keV and the temperature of the modified blackbody for \sgrb bursts. The results from observation 00340573000 are presented with open symbols, while those from observation 00340573001 are in solid ones. The spectral results of SS, SL, DSL+DLS, MS and MO bursts are shown as circles, triangles, squares, upside down triangles and crosses, respectively. The solid line is the best fit power law using spectral results of all types of bursts from \sgrbnos.  \label{fig-ktfluxb}}
\end{figure}

We also investigated the correlative trend between the average unabsorbed flux and the temperature of modified blackbody for all stacked spectra (Figure \ref{fig-ktfluxb}). In general, these two quantities are well correlated (the Spearman's rank correlation coefficient is 0.839 and the chance occurrence probability is $4.9\times10^{-5}$). This correlation can also be described with a single power law trend, having an index of $2.1\pm0.4$ ($\chi^2/dof=23.82/14$), which agrees with that from the fit using only short blocks/bursts from \sgrnos. 

The single blackbody (BB) function has also been used to describe the spectral shape of \sgrb bursts in the 8-200 keV band \citep{vonkienlin2012}. To test this model, we fit the stacked spectra with a BB in a similar manner as to the MBB+RCS case. First, we fit all types of burst spectra in both observations simultaneously, and obtained the galactic absorption of $(2.6\pm0.2)\times10^{22}$\,cm$^{-2}$ ($\chi^2/dof=447.01/466$) and $(2.1\pm0.3)\times10^{22}$\,cm$^{-2}$ ($\chi^2/dof=91.48/111$). Then, we fit each stacked spectrum with the $n_{\rm H}$ set at their corresponding values. From the statistical point of view, the BB model fits to the stacked burst spectra are acceptable. However, the $n_{\rm H}$ values obtained are much lower than those reported earlier in the literatures, as well as those obtained with the MBB+RCS model fit. Moreover, even if these lower $n_{\rm H}$ are considered as the case, the resulting total $\chi^2$ value from the BB model fits to all stacked spectra is larger by 11.7 than that from the MBB+RCS model, which has the same number of free parameters. Therefore, we concluded that the MBB+RCS model is preferred over the BB model for the stacked burst spectra of \sgrb in the $0.3-10$\,keV band. 

In order to compare the shape of the stacked spectra with that of the persistent source, we fit them with the STEMS model likewise, as the sum of a STEMS and a single power law function which successfully described the persistent spectra in the same burst active episode \citep{ng2011}. None of our stacked spectra can be fit with STEMS model or even with an additional power law. 

\section{Differential fluence distribution}

\subsection{\sgr}
\begin{figure}[ht]
    \includegraphics[scale=0.8,bb=20 220 500 600]{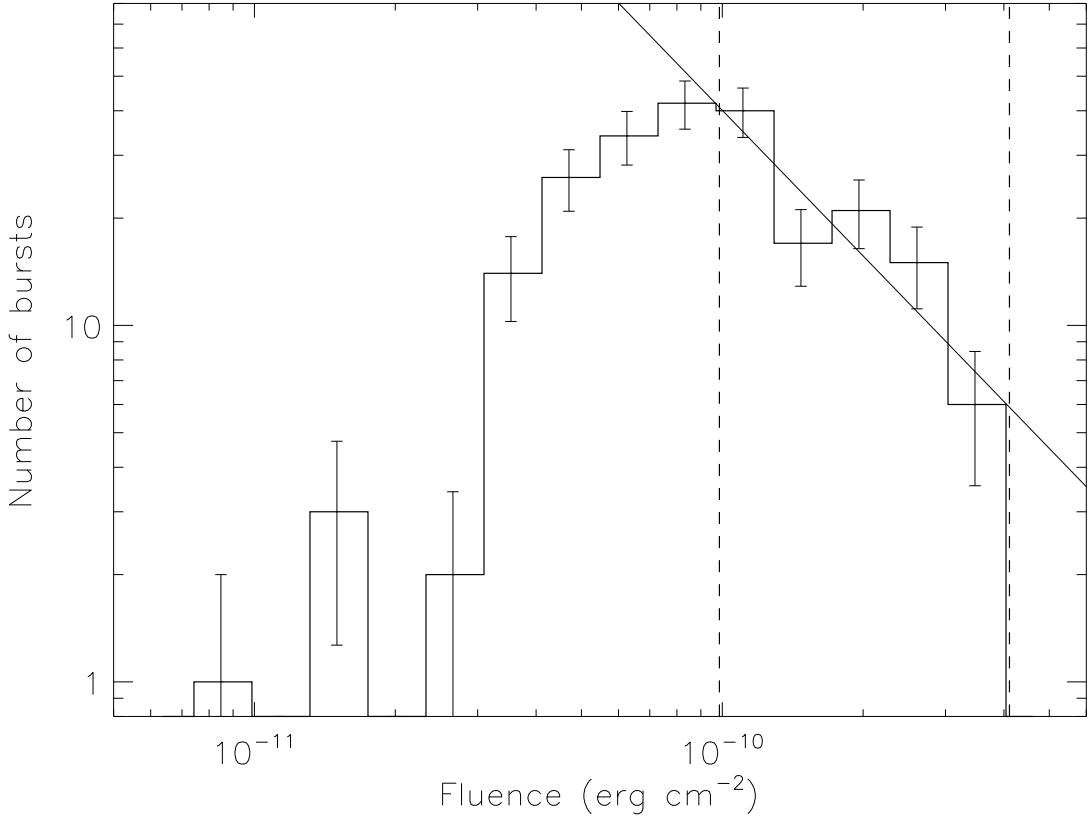}
\caption{The differential fluence distribution of 221 short bursts from \sgrnos. The solid line presents the best fit power law model to the histograms within the reliable burst detection fluence range (see the text) shown with vertical dashed lines. \label{fig-lgnlgs}}
\end{figure}

To study the differential fluence distribution of \sgrnos, we focused on SS \& MS bursts and short blocks of DSL \& DLS bursts since these types have been shown to exhibit distinct burst nature (Section 5.1). We estimated the fluence of each burst using a conversion factor, which is the average of the total energy to count ratio for the selected groups of bursts (listed in the rightmost column of Table 2). The average conversion factor of the stacked spectra of short bursts from \sgr is $8.2\times 10^{-12}$\,erg\, count$^{-1}$. In Figure \ref{fig-lgnlgs}, we exhibit the differential fluence distribution of 221 bursts in equal logarithmic steps. 

To further study the fluence distribution of \sgr bursts, we first examined the instrumental effects, that might limit the detection sensitivity. The high fluence cutoff is determined by the pile-up level. For this \textit{XMM-Newton} observation, we excluded piled-up bursts which have more than 50 counts in a second. All 221 short bursts in our distribution are below this limit, that corresponds to an energy fluence of $4.1 \times 10^{-10}$\,erg\,cm$^{-2}$. We determined the lower fluence cutoff through numerical simulations as follows: We assumed that a flat differential fluence distribution and the burst counts are evenly distributed throughout the burst interval. We performed simulations on a series of total burst counts ranging from 2 to 50. For each total count value examined, we iterated the following steps for 10000 times: First, we randomly selected a burst duration drawn from the distribution of short bursts in the right panel of Figure \ref{fig-durdis}. Then, we generated a background included event list by adding a 20 s long segment of the observed persistent emission of \sgr with the simulated burst events with the specified duration. These two event lists were aligned at their mid-times. Finally, we applied the Bayesian blocks representation process to the simulated event list and determined whether the simulated burst can be detected or not. We found from these simulations that the detection probability of bursts with at least 12 total counts (corresponding to $9.9\times 10^{-11}$\,erg\,cm$^{-2}$) is higher than 99\%. Therefore, we set the lower limit at this fluence level. 

We have 97 short bursts in the fluence range where the instrument as well as our burst identification process have the complete detection capabilities. We fit the differential burst fluences in this fluence range with a single power law, and determined the index as $-2.4\pm0.3$ ($\chi^2/dof=5.9/3$). In order to avoid the binning effects, we also applied the maximum likelihood estimation to determine the power law index from the unbinned burst fluences and obtained $-2.3\pm0.3$ with $1\sigma$ errors. 

\subsection{\sgrbnos} 

\begin{figure}[ht]
\includegraphics[scale=0.8,bb=20 220 500 600]{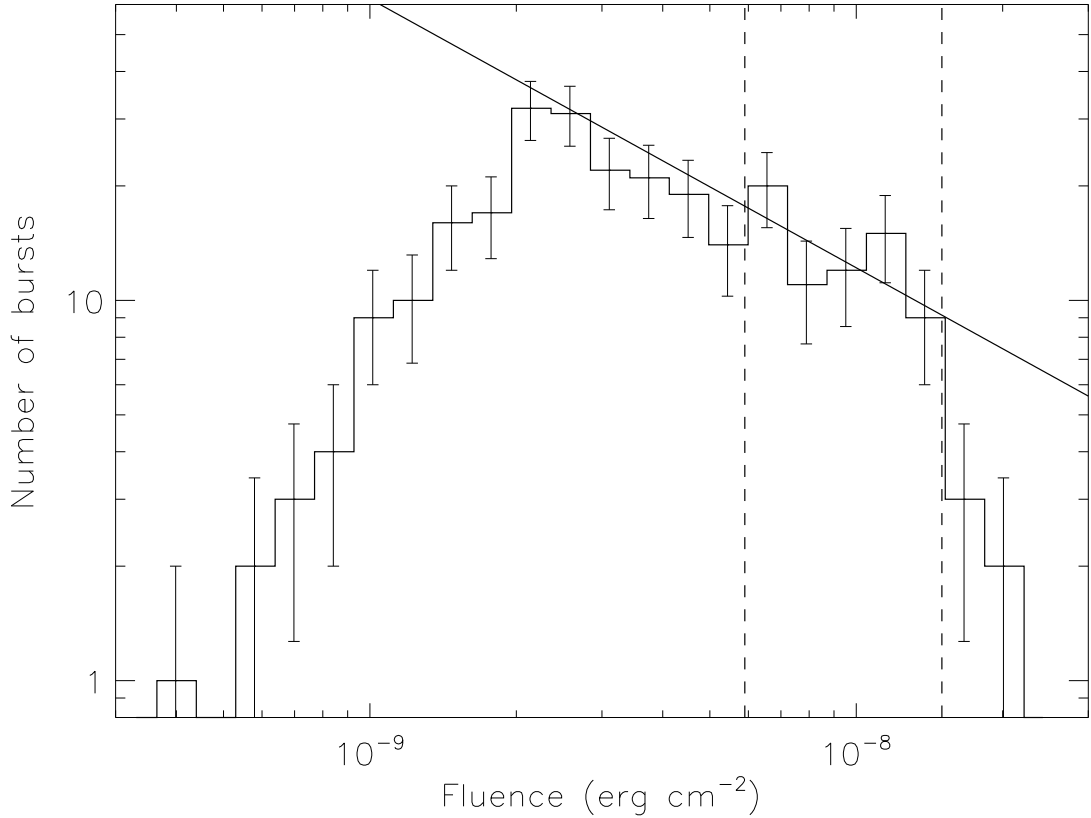}
\caption{The differential fluence distribution of 273 bursts of \sgrbnos. The vertical dashed lines are the fluence limits in between which the burst detection number is reliable. The solid line presents the best fit power law model to 5 reliable fluence bins.  \label{fig-lgnlgsb}}
\end{figure}

We studied the burst fluence distribution of \sgrb using 273 events from all types of the Bayesian blocks profiles, as they entirely clearly represent burst origin from spectral investigations (Section 5.2). Here we already excluded the only two bursts from the forth orbit of the observation 00340573000, which suffered from a much higher persistent flux ($\sim64.2$\,counts\,s$^{-1}$) than the other orbits. Similar to \sgrnos, we calculated the fluence of each burst by converting the total burst counts into corresponding energy. If a burst has been grouped into any stacked spectrum, then we used the spectral fit results to determine the conversion factor (see the last column of Table \ref{tab-bstspec}). Otherwise, we adopted the average value of the conversion factors in its orbit or observation, that is, $1.4\times10^{-10}$\,erg\,count$^{-1}$, $1.5\times10^{-10}$\,erg\,count$^{-1}$, and $1.4\times10^{-10}$\,erg\,count$^{-1}$ for orbits 1, 2 and 3 of observation 00340573000, and $1.4\times10^{-10}$\,erg\,count$^{-1}$ for observation 00340573001. In Figure \ref{fig-lgnlgsb}, we present the size distribution of these 273 bursts. 

Following the same procedure for \sgrnos, we also calculated the fluence limits of  bursts with detection rate over $99\%$ for \sgrbnos. The higher fluence limit is $1.5\times10^{-8}$\,erg\,cm$^{-2}$, corresponding to the energy fluence of 100 counts (i.e., the pile up limit) using the average conversion factors derived from all investigated orbits. Due to the significant change of the persistent flux level, we performed four sets of simulation for the first three orbits in observation 00340573000 and the observation 00340573001. We obtained the minimum burst count with over 99\% detection rate as 42, 28, 39, and 33, respectively. After converting them into the energy fluence, we adopted the highest value of $5.9\times10^{-9}$\,erg\,cm$^{-2}$ as the lower fluence limit. Therefore, we selected 68 bursts with fluence in between these two limits. We fit the distribution within this range with a single power law function and obtained the best fit index as $-1.7\pm0.5$ ($\chi^2/dof=2.81/3$). We also calculated the power law index using the maximum likelihood method, which yields the most probable index with $1\sigma$ confidence range as $-1.7\pm0.5$.

\section{Discussion}


In this paper, we introduced for the first time the Bayesian blocks method to search for weak magnetar bursts in X-ray observations using time tagged event data. We successfully identified 320 bursts from \sgr observed with \textit{XMM-Newton} and 404 bursts from \sgrb in two \textit{Swift}/XRT observations. The Bayesian blocks burst search process takes advantage of fine temporal resolution of the instruments, without running into any limitations introduced by binning the data. The Bayesian blocks is effective in a wide range of burst energetics, especially in identifying weak bursts from the fluctuations of the persistent source emission. The unabsorbed fluence in the $0.5-10$\,keV band of bursts identified with this method ranges from $7.4\times10^{-12}-3.9\times10^{-10}$\,erg\,cm$^{-2}$ for \sgrnos, and $4.0\times10^{-10}-1.9\times10^{-8}$\,erg\,cm$^{-2}$ for \sgrbnos. In general, these bursts are dimmer by about two orders of magnitude than those had been extensively analyzed in literatures: $\sim10^{-8}$\,erg\,cm$^{-2}$ in $8-200$\,keV \citep{lin2011} and $\sim10^{-10}$\,erg\,cm$^{-2}$ in $2-40$\,keV \citep{nakagawa2011} of \sgr bursts; $10^{-8}$\,erg\,cm$^{-2}$ in $0.3-300$\,keV \citep{enoto2012} and $>10^{-8}$\,erg\,cm$^{-2}$ in $8-200$\,keV \citep{vdh2012,vonkienlin2012} of bursts from \sgrbnos. We note that these fluence values were accumulated in different energy ranges. However, the broadband spectral analysis from 0.5\,keV to 200\,keV showed that the fluence in $8-200$\,keV energy range is only about an order of magnitude larger than that in the $0.5-10$\,keV band \citep{lin2012a}. Therefore, after calibrating for the energy ranges, our samples from the Bayesian blocks burst search technique still represent the dimmest burst population from magnetars. 


The Bayesian blocks representation allows a direct way to investigate the burst temporal morphology. A large number of burst counts accumulated into a single block, based on statistics, remarkably simplifies the burst lightcurve. Unlike the burst search performed with the binned data, this method can also reveal statistically significant temporal structures, which would otherwise be smoothed out in the binned lightcurve. In our sample for both magnetars 70.95\% of burst profiles are represented with a single short block (SS); 5.96\% of bursts have more than one but all short blocks (MS); 14.71\% of events show a stable flux for longer than half a second (SL); 5.59\% of burst profiles have two blocks, a short block is either followed or preceded by a lower flux structure (DSL \& DLS); and the remaining 2.79\% of events exhibit more than one short peak or long smooth emission profile (MO). We show that except for the second orbit of observation 00340573000 which covered the burst storm of \sgrbnos, all other observations and orbits exhibit comparable percentage of different burst types (see Table \ref{tab-bbtype}). 

Based on the Bayesian blocks profiles, we can also directly determine the duration of a burst from the block size, without running into difficulties of the background modeling, that can be severe for dim bursts. We can, therefore, study statistical behavior of the temporal properties of the dimmest magnetar bursts. The bursts duration distribution of \sgr in 0.2-10 keV follows a double-Gaussian trend with two components dominated by short and long bursts. The mean value of the duration of the short bursts as obtained with a Gaussian function fit is $\sim87$\,ms. This is shorter than the average $T_{90}$ duration of \sgr bursts, while it is comparable to their average emission times ($\tau_{90}$), which is the duration needed to collect 90\% of the burst total counts regardless the order of their arrival times, in 8-200\,keV energy range \citep{lin2011}. For typical single pulse bursts, the average $T_{90}$ duration is approximately equal to the average $\tau_{90}$ of all bursts \citep{gogus2001}. It is, therefore, expected that the distribution of our short blocks sample of \sgr to be consistent with the $T_{90}$ duration distribution of bursts with simple profile. For \sgrbnos, the distribution of all events follows a single Gaussian-like trend, similar to that of more energetic bursts seen with \textit{Fermi}/GBM \citep{vdh2012,vonkienlin2012}. These agreements support the burst nature of short blocks from \sgr and all events from \sgrbnos. 


We performed detailed spectral analysis of the weakest magnetar bursts grouped into eight stacked spectra for \sgr and sixteen for \sgrb based on their temporal profiles and fluences. All the stacked spectra can be fit adequately with the MBB+RCS model. To investigate the connection between the burst and persistent emission, we also studied the accumulated burst spectra with the STEMS model, that is developed to better understand the persistent emission spectra of magnetars. We found that the STEMS model can fit only the stacked spectra of SL and long blocks of DSL \& DLS events of \sgr (54 out of 320 bursts). As we discussed above, these types of bursts accounted for an additional Gaussian component in the combined duration distribution of \sgr bursts. Based on these, we cannot conclusively determine whether these weak and long blocks of \sgr are indeed representing bursts or rapid enhancements of the persistent source emission. Stacked spectra of all other types of \sgr bursts and entire \sgrb events cannot be fit with the STEMS model, and therefore, they likely represent a distinct origin other than the persistent emission. As a consequence, we strengthened the conclusion of \citet{lin2012b} with much dimmer magnetar bursts that the emergent spectra of the burst and persistent emission are characteristically different, even though their origin likely involve an underlying extremely strong magnetic field.

We found that \sgrb bursts detected with XRT have, on average, higher modified blackbody temperature and unabsorbed flux than those of \sgr observed with \textit{XMM-Newton}. However, we obtained a correlation between the flux and temperature for both sources. In particular, power law trends with indices of $2.5\pm1.0$ and $2.1\pm0.4$ are followed between these two quantities using only spectrally confirmed bursts of \sgr and \sgrbnos, respectively. Note that both power law indices have large deviations from the expectation of a pure blackbody model (i.e., 4), which may indicate the signature of the strong magnetic fields on the emergent burst emission. We also found that for \sgr, the surface temperature and flux of the persistent emission (obtained with the STEMS model) cannot be inferred from the extrapolation of the power law trend of neither all burst types, nor only the short burst blocks. It is important to note that the temperatures from STEMS and MBB+RCS models arise from different physical settings: the former is temperature of the thermal emission from the surface of the neutron star and the latter is the temperature of the thermalized bubble emission modified by the strong magnetic field.

\begin{figure}[h]
\includegraphics[scale=0.8,bb=10 360 80 720]{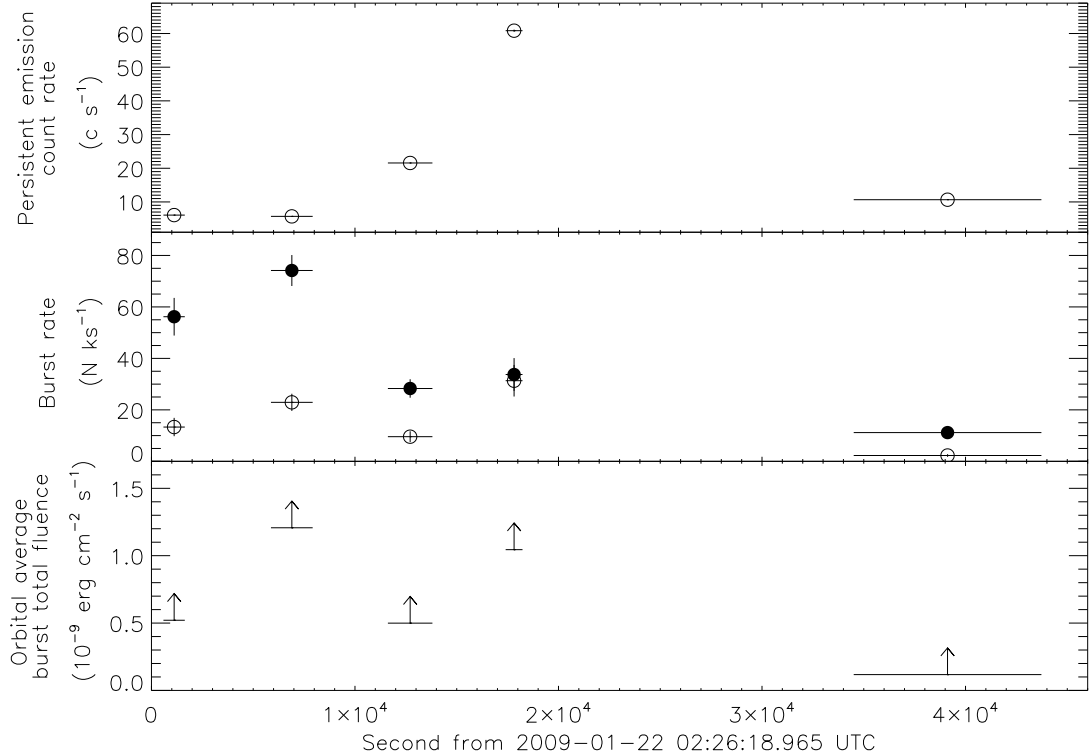}
\caption{Evolutions of the persistent emission count flux (\textit{top}), burst rate (\textit{middle}) and averaged total burst unabsorbed fluence in $0.5-10$\,keV over the exposure time (\textit{bottom}) of \sgrb in two \textit{Swift}/XRT observations on January 22, 2009. The first observation was shown as four orbits separately. The solid dots in the middle panel are the burst rate including all events, while the open circles are the rate of only piled-up bursts. The upper arrows in the bottom panel indicate the lower limit of the averaged fluence due to the piled-up bursts. \label{fig-bstperlead}}
\end{figure}

The persistent emission flux of \sgrb changed dramatically during its burst active episode in 2009 \citep{scholz2011}. The X-ray flux of the source reached to its peak in 2009 January 22, the most burst active day. Using the burst sample we identified with the Bayesian blocks method, we compared the evolution of the persistent flux in conjunction with those of the burst number and the orbital averaged burst energy fluence (see Figure \ref{fig-bstperlead}). We find that the burst rate peaks in the second XRT orbit, that is about 11 ks prior to the peak of the persistent flux (count rate) in the fourth orbit.
This behavior indicates that the sudden release of a large amount of burst energy may ignite the heating up of the neutron star crust, and result with the increase of the temperature and flux of the persistent emission, which were observed by \citet{scholz2011}. We note the important fact that the rate of strong (piled-up) bursts also peaked in the fourth XRT orbit. Therefore, it may also be possible that the strong bursts in the fourth orbit may have driven simultaneous persistent flux increase, rather than gradual surface heating.

We studied the fluence distribution of \sgr bursts using the dimmest sample in the 0.5$-$10 keV band for the first time. The distribution follows a power law trend as $dN \propto S^{-2.3}dS$. \citet{lin2011} formed the fluence distribution of \sgr bursts using 29 events seen with GBM in the 8$-$200 keV and obtained a power law index of $-1.5$, which is much flatter than the trend at the lower energy level. This may indicate the fact that there is a break somewhere between the fluences of 4$\times$10$^{-10}$ and 10$^{-7}$ erg cm$^{-2}$. We have further investigated this possibility as follows: There were ten bursts detected from \sgr with GBM during the course of \textit{XMM-Newton} observation \citep[bursts $\#5-14$ in Table 1 of][]{lin2011}. Note the fact that Fermi is in a low altitude orbit, therefore, the source was blocked by the Earth by about 50\% of the exposure of \textit{XMM-Newton}. Assuming that the source was emitting bursts at a constant rate (which is implied from the \textit{XMM-Newton} observation), we considered that about 20 bursts must have been detected with GBM over the time that 320 bursts were recorded with \textit{XMM-Newton}. We then calibrated our fluence distribution using the ratio of the number of bursts detected by two instruments, and extrapolated the distribution to the higher fluence regime to estimate the number of bursts expected if there is no break (namely, the index of $-2.3$ spans over all fluences). We found that the expected number of bursts from power law distribution with index of $-2.3$ is about two orders of magnitude lower than the number of bursts detected with GBM. This result strengthened the possibility that there might indeed be a break in the broad fluence distribution \sgr.

\citet{scholz2011} constructed the fluence distribution of \sgrb bursts in the 0.5$-$10 keV energy band and obtained a power law index of $-1.6$, that is in agreement with our investigations here ($-1.7\pm0.5$). A similar power law index was reported for the fluence distribution of \sgrb bursts within the fluences of $1.8\times10^{-7} - 3.2\times10^{-6}$\,erg\,cm$^{-2}$ \citep{vdh2012}. Note that the upper fluence limit of our burst sample is nearly equal to the detection limit of GBM bursts. Therefore, there is clearly no evidence for any break in the fluence distribution of \sgrb over three orders of magnitude in fluence.


According to the magnetar model, the typical short bursts result from the release of crustal energy, triggered by the cracking of the neutron star crust by the strong magnetic field \citep{dt92,td95}. The required crustal magnetic field strength can be estimated from the energy of a burst, the length of the crustal crack and the upper limit of the lattice strain, that is typically 10$^{-3}$ \citep{td95}. Moreover, one could also estimate the displacement of the magnetic footprints and the minimum excitation radius of the Alfven wave by using the crack length and the magnetic field strength \citep{td95}. Our earlier STEMS model fit to the persistent emission of \sgr resulted in an average surface magnetic field strength of $2.2\times10^{14}$\,G for. Using this value, we obtain a crack length scale of $\sim10^{-1}$\,m, for the weakest bursts from this magnetar (that is, SS bursts with SNR $\le3$). We estimate the corresponding displacement of the magnetic footprints as $\sim2\times10^{-2}$\,m and the minimum Alfven excitation radius as $\sim5$\,m. Note the fact that the surface magnetic field topology of the neutron star is not necessarily uniform and the field strength may exceed the average value under more localized settings. Therefore, our estimates for the length scales are the lowest limits for \sgr bursts.

\acknowledgments 

L.L. is funded through the Post-Doctoral Research Fellowship of the Turkish Academy of Sciences (T\"UBA).

\end{document}